\documentclass[aps,prb,twocolumn,superscriptaddress,noshowpacs,a4paper,10pt]{revtex4}
\pdfoutput=1
\usepackage{graphicx}

\begin{document}

\author{\firstname{Roberto} \surname{Guerra}}
\email{guerra@sissa.it}
\affiliation{International School for Advanced Studies (SISSA), Via Bonomea 265, I-34136 Trieste, Italy.}

\author{\firstname{Stefano} \surname{Ossicini}}
\affiliation{Centro Interdipartimentale En\&Tech and Dipartimento di Scienze e Metodi dell'Ingegneria, Universit\`a degli Studi di Modena e Reggio Emilia, via Amendola 2 Pad.\ Morselli - 42122 Reggio Emilia, Italy.}

\title{Preferential positioning of dopants and co-dopants\\in embedded and freestanding Si nanocrystals}

\begin{abstract}
  In this work we aim at understanding the effect of n-type and p-type substitutional doping in the case of matrix-embedded and freestanding Si nanocrystals. By means of {\it ab initio} calculations we identify the preferential positioning of the dopants and its effect on the structural properties with respect to the undoped case. Subsequently, we consider the case of phosphorus and boron co-doped nanocrystals showing that, against the single-doping situation, the energetics strongly favors the binding of the impurities at the NC surface. Indeed we demonstrate that the polar B-P bond forms a stable permanent electric dipole that radially points inwards the nanocrystal. Such noteworthy characteristic and its physical consequences are discussed alongside new potential applications.
\end{abstract}

\maketitle

\vspace{0.5cm}
\section{Introduction}\label{sec.intro}
\noindent The use of silicon nanocrystals (Si-NCs), embedded in wide-gap materials, or in colloidal solution, has found large employment in applications such as optoelectronics and photonics,\cite{daldosso2009,blas2009,cheng2011,mastronardi2011} photovoltaics, \cite{conibeer2012,beard2007,Liu2009,gregor2012,ivan2012,loper2013} and bio-imaging.\cite{erogbogbo2008,erogbogbo2011,fujii2012}
Several advancements in the fabrication techniques permit nowadays to control with good precision the size, shape, and density of the Si-NCs. Colloidal Si-NCs allows, in principle, the fabrication of large-area optoelectronic devices by vacuum-free printable processes. Usually, in order to avoid NCs agglomeration one functionalizes the nanocrystals surface with organic molecules, but this can be detrimental to the electron transport properties.\cite{law2010} Thus, the production of colloidal Si-NCs without surface functionalization processes is highly desirable. This has been recently obtained by synthesizing all-inorganic Si-NCs using simultaneous doping with n- and p-type impurities.\cite{fujii2012,fukuda2011,fujii2013a,fujii2013b}
Regarding matrix-embedded Si-NCs, they offers advantages in terms of stability, low-cost manufacturability and the development of CMOS compatible devices. However, the high resistivity of the embedding matrix reduce the possibility to employ such composite material in applications that require a high carrier mobility. One way to enhance the current of Si-NCs-based devices is to increase the density of the NCs in the samples, thus increasing the silicon content and reducing the tunneling distance between the Si-NCs.
The downside of this approach lies in the difficulty of controlling the density of the Si-NCs distinctly from other parameters (e.g. size, shape, and so on), and in the limited efficiency of quantum confinement while approaching the percolation threshold, thus affecting the optical properties.\cite{balberg}
In alternative, the possibility of introducing dopants has been explored in recent years in order to improve transport and other fundamental characteristics.\cite{khria2012,gutsch2012,mathiot2012} Clearly, in the case of SiO$_2$-embedded Si-NCs one may focus on doping the sole SiO$_2$ matrix, thus improving the conductivity of the sample while not altering the optical absorption of the Si-NCs, the latter tunable for example by changing the NCs size distribution.\cite{JAP2013}
\\Differently than bulk materials, the doping of nanostructures involves additional complications arising from the presence of interfaces, and of materials with different affinities in hosting the dopant species.\cite{Norris} In practice, experimental samples of embedded Si-NCs present inhomogeneous impurity concentrations, that depend on the fabrication technique.\cite{khelifi2013} In other words, since dopants tend to diffuse during the fabrication process, one cannot simply limit the doping to a selected region of the material.
\\Therefore, in order to correlate the effect of the dopants to the observed quantities, it is important to understand the preferential positioning of the impurities in the nanostructures. In this context, the presence of defects in unpassivated samples will likely favor the presence of dopants in the proximity of interfaces (i.e. interstitial doping).\cite{khelifi2013} In practice, at low impurity concentrations the impurity atoms will act as passivants, correspondingly improving the optical yield. Conversely, in the case of high concentrations or of well passivated (or undefective) samples, the impurities substitute to the host material atoms, by diffusing towards the most energetically favorable sites.\cite{gutsch2012}
\\In the following we investigate the effect of n-type and p-type substitutional doping in the case of Si-NCs embedded in SiO$_2$ and of freestanding, colloidal Si-NCs. By means of {\it ab initio} total-energy calculations we identify the preferential location of dopants in the nanostructure, and its effect on the structural properties with respect to the undoped case. Finally, the case of co-doped NCs will be investigated and discussed, in particular for colloidal NCs, alongside its potential applications.

\section{Structures and Method}\label{sec.structuresandmethod}
\begin{figure*}[t!]
  \centering
  \includegraphics[draft=false,width=\textwidth]{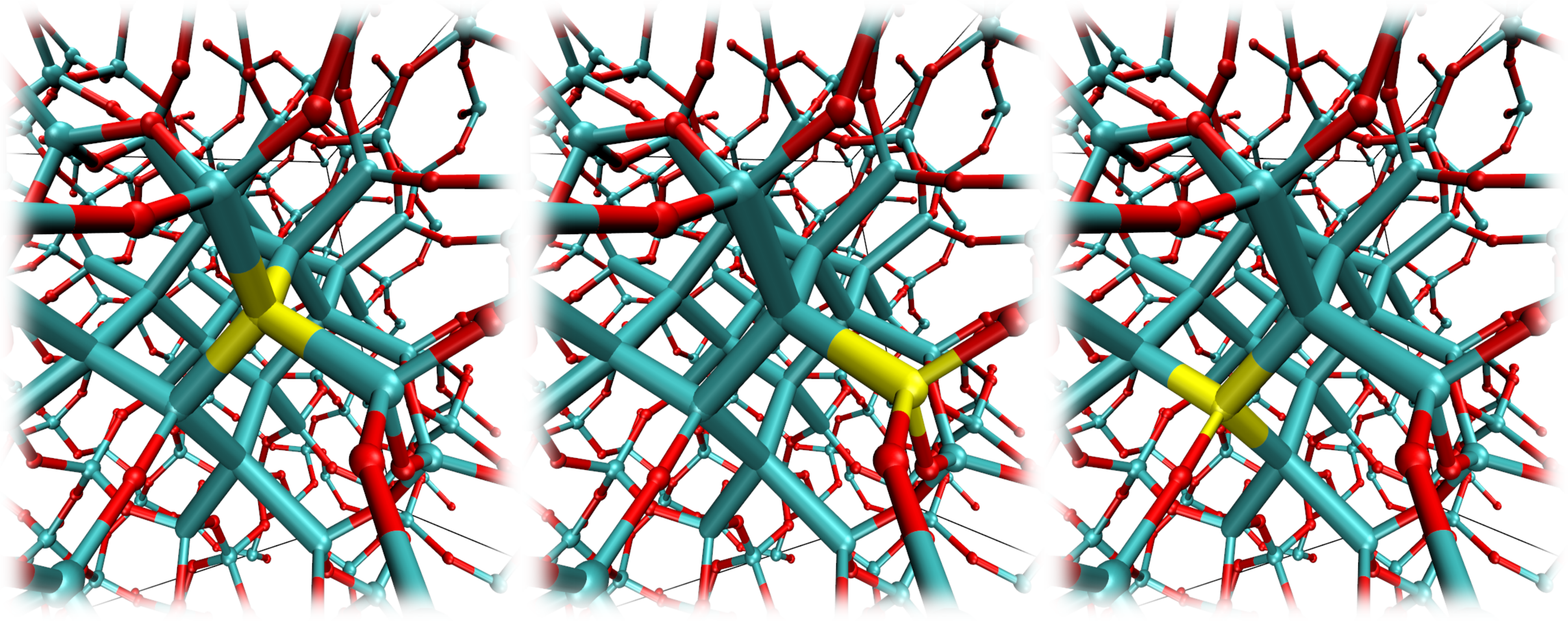}
  \caption{\small Ball-and-stick representation of a SiO$_2$-embedded Si-NC with an impurity atom placed inside the NC (left panel) and at two different interfacial sites, corresponding to Si$^{3+}$ (central panel) and Si$^{1+}$ (right panel) in the undoped structure. Red and cyan spheres represent O and Si atoms, respectively. Si-NC atoms are connected by thick sticks, while the yellow region marks the position of the dopant. }\label{fig.structures}
\end{figure*}

All the structural and electronic properties have been obtained by full \textit{ab-initio} calculations in the framework of density functional theory (DFT) using the {\small SIESTA} package.\cite{siesta1,siesta2} Calculations have been performed using norm-conserving pseudopotentials within the local-density approximation (LDA).
An energy cutoff of 250\,Ry on the electron density and no additional external pressure or stress have been applied. Consistently with convergency trial runs, all the calculations have been performed with a Gamma point k-sampling and standard double-$\zeta$ basis set for all the atoms. Atomic positions and cell parameters have been left totally free to move.
\\The SiO$_2$-embedded NCs have been generated starting from a 3$\times$3$\times$3 betacristobalite-SiO$_2$ matrix, by removing all the oxygen atoms inside a sphere whose radius determines the NC size. The so-obtained Si-NC, embedded in the SiO$_2$, presents perfectly coordinated atoms and the same Si-Si distance of betacristobalite, corresponding to about 3.2\,\AA. After the relaxation, the Si/SiO$_2$ interface forms strongly stressed bonds, while far from the interface the bulk atomic densities are recovered.\cite{PRB2}
\\We have introduced the dopant impurity in a substitutional Si-site located in the NC center, at the Si/SiO$_2$ interface, or in the SiO$_2$ far from the NC. Since the interfacial Si atoms can neighbour with one-to-three oxygen atoms (i.e. Si$^{1+}$, Si$^{2+}$, Si$^{3+}$), for each structure we have considered all the possible oxidation type for the interfacial doping.\cite{kim2007} Each structure has been fully relaxed after doping in order to include the effect of the structural rearrangements around the dopant atom. Examples of so-obtained doped structures are shown in Figure~\ref{fig.structures}.
\\All the calculations have been performed for the Si$_{32}$, the Si$_{47}$, and the Si$_{35}$ NCs embedded in SiO$_2$. However, since we have obtained equivalent results for the three different cases, for the sake of simplicity we will present only the results of the latter system in the following.
\\The freestanding colloidal Si-NCs have been obtained from bulk silicon by a spherical cutoff, and by passivating all the dangling bonds with OH groups (oxidized Si-NCs).
The choice of OH as Si-NCs termination has several reasons. First of all, we have already shown\cite{PRB2,PRB3} that, beside strain effects, there is a strong similarity between the calculated electronic and optical properties for the silica embedded Si-NCs and for the Si-NCs simply passivated by the OH groups. Thus these properties are almost completely determined by the barrier provided by the first shell of oxygen atoms.
Moreover it is worth to note that, from the present and other\cite{carvalho,carvalho1} calculations, the energetics of the doping sites seems crucially affected only by the termination type while the presence of an embedding matrix should play a minor role.
Like for the embedded Si-NCs, in our calculations each freestanding structure has been fully relaxed after doping. We have used a 3\,nm-sized periodic cubic cell for the freestanding NCs smaller than the Si$_{87}$ (diameter\,$\simeq$\,1.36\,nm), and 5\,nm for the others. The error on the total dipole due to the interaction with the NC replica has been estimated by applying, for the largest NC, the Martyna-Tuckerman correction to the scf potential in order to mimick the case of a truly insulated NC.\cite{mt} The so-obtained total dipole was about 6\% smaller than the uncorrected one; this value should indicate the maximum error on the calculated dipole moments.
\\In the following we will consider the Si$_{87}$-(OH)$_{76}$ as reference system for the study of co-doping.

\section{Single-doped Silicon Nanocrystals}\label{sec.single-doped}
\begin{figure}[b!]
  \centering
  \includegraphics[draft=false,width=\columnwidth]{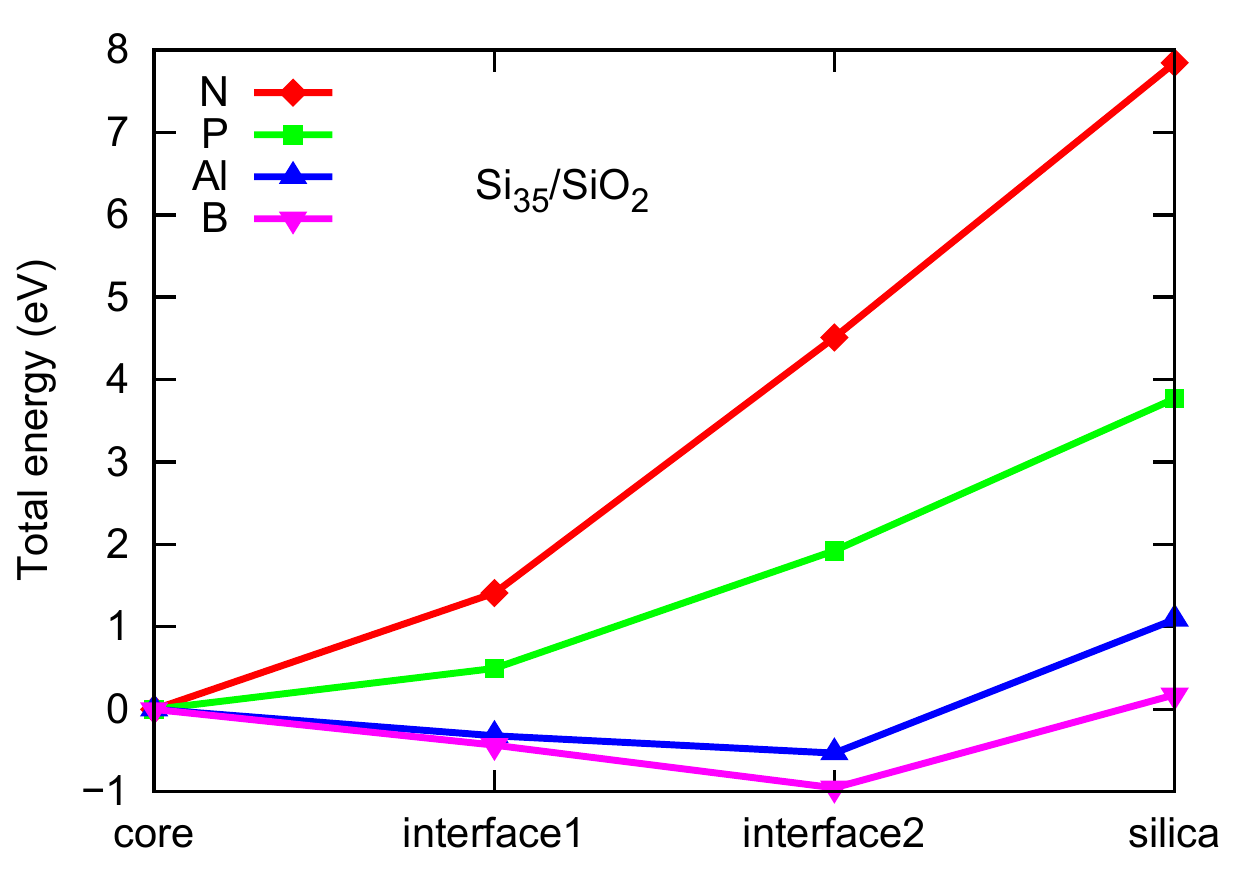}
\caption{\small Total energy of the Si$_{35}$ NC embedded in the SiO$_2$ structure and doped with N, P, Al, or B, substitutional of a Si atom in the NC center (core), at the interface bonded to one or two oxygens (interface1, interface2), or in the SiO$_2$ far from the NC (silica). For a better comprehensibility each line has been shifted so that the structure with the dopant in the NC center has zero energy. }\label{fig.totenergy}
\end{figure}

We report in Figure~\ref{fig.totenergy} the total energy comparison for the Si$_{35}$ NC (diameter $\simeq 1$\,nm) embedded in SiO$_2$, single-doped with a group-V (N or P) or a group-III (Al or B) atom placed at different substitutional sites.
The choice of taking into account N impurities is related to the fact that N presents very different electronegativity, oxidation states, and electron affinity, with respect to P. Conversely, As atoms should behave similarly to P, given an almost exact matching of the above characteristics.

\begin{figure}[t!]
  \centering
  \includegraphics[draft=false,width=4.cm]{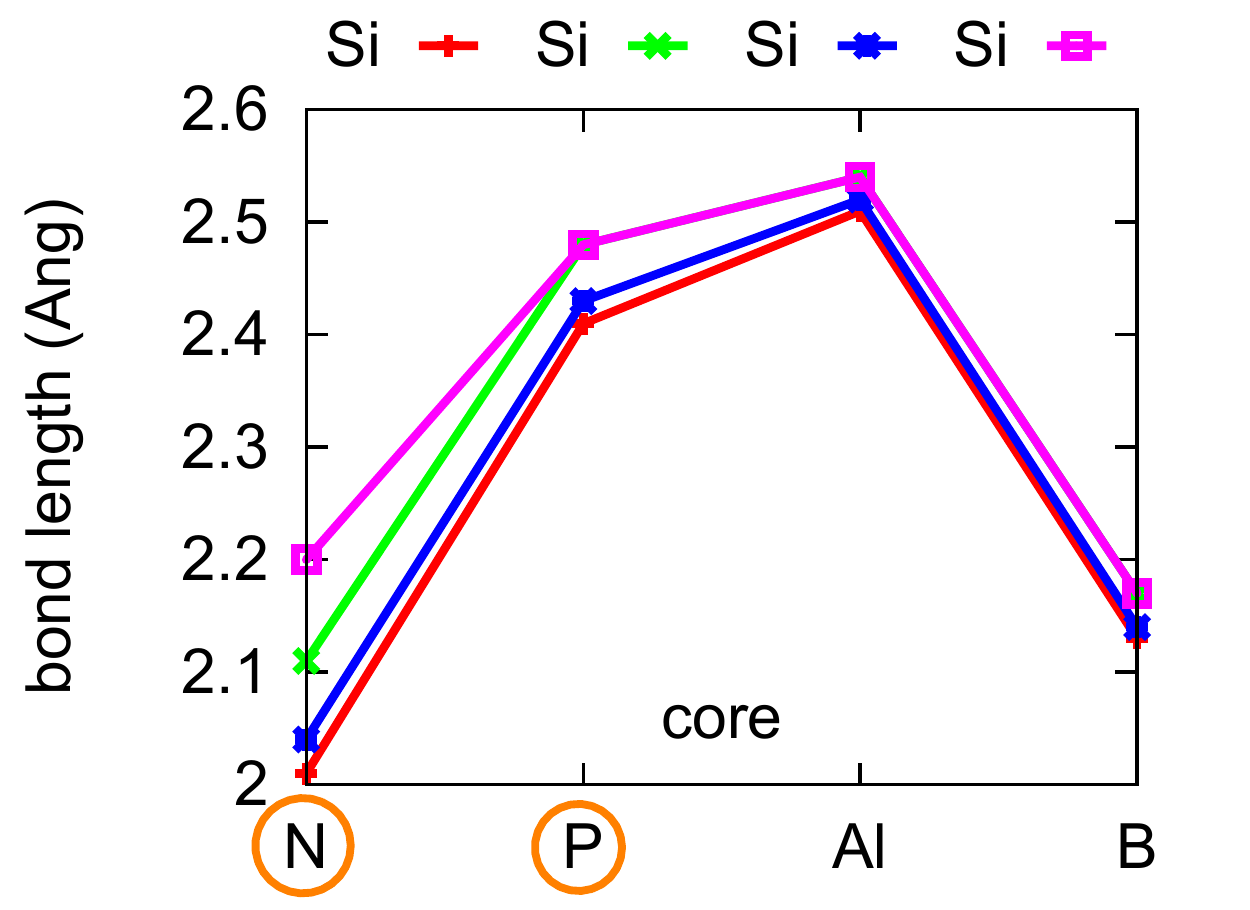}
  \includegraphics[draft=false,width=4.cm]{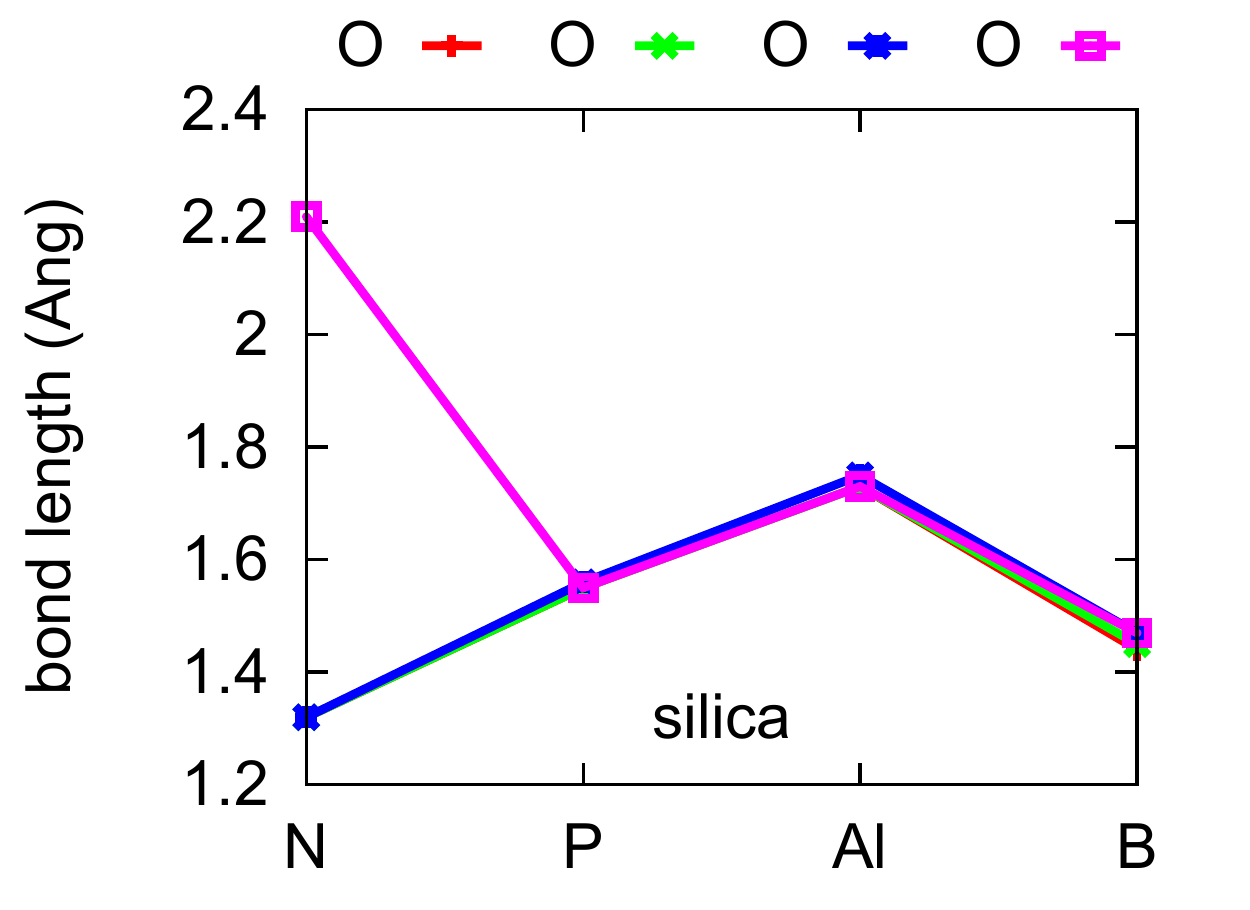}\\
  \includegraphics[draft=false,width=4.cm]{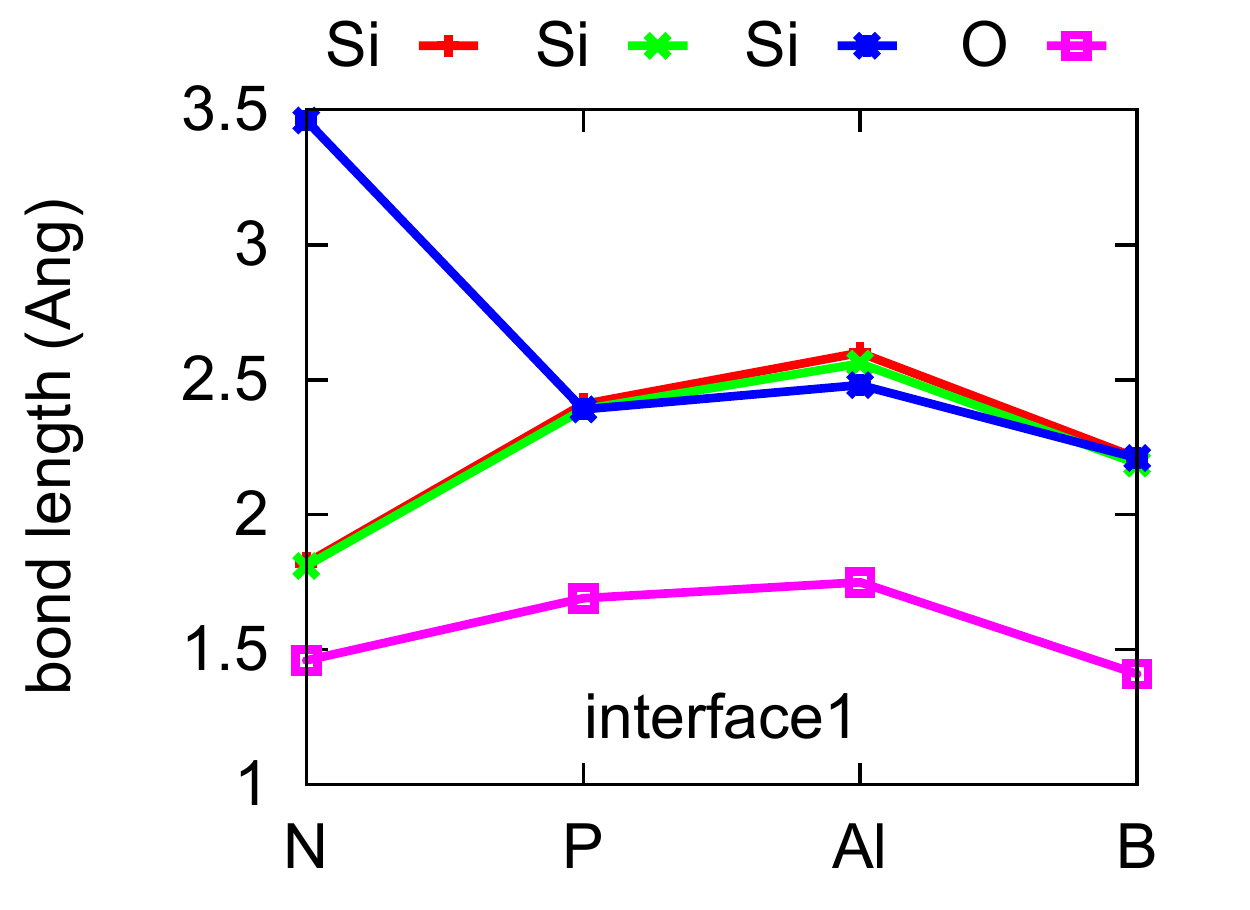}
  \includegraphics[draft=false,width=4.cm]{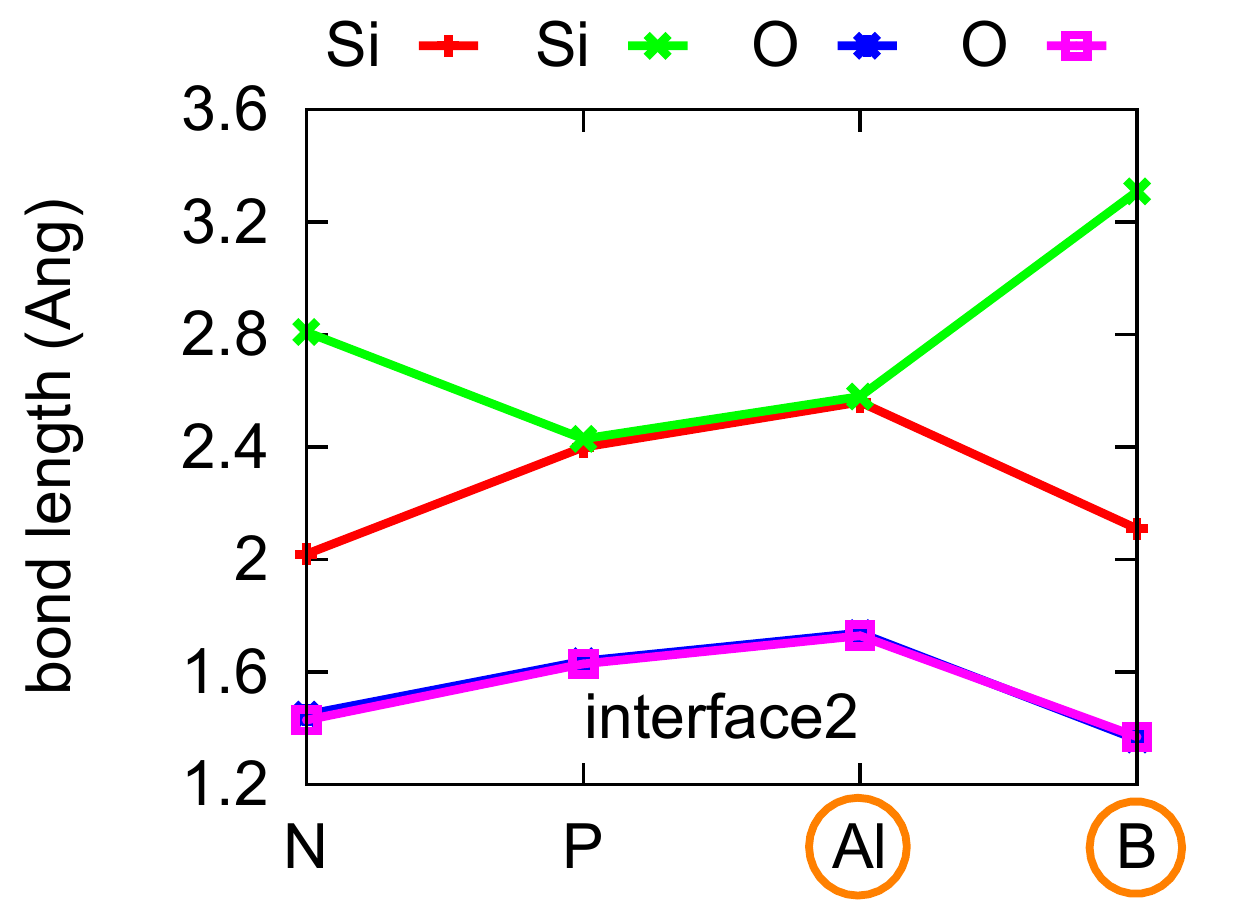}
\caption{\small Length of the bonds formed by the dopant with its neighbours, for different dopant site in the Si$_{35}$ NC embedded in SiO$_2$ (please refer to Fig.~\ref{fig.totenergy} for nomenclature). Circles highlight the most stable site for each dopant.}\label{fig.bond_lengths}
\end{figure}

In Figure~\ref{fig.totenergy} the minimum of each curve identifies the energetically favored site of the corresponding dopant.
The data clearly indicate that for n-type doping (P and N) the impurity tends to settle in the NC core, while for p-type doping (Al and B) the interfacial sites are favored. These results are on line with recent theoretical calculations made on oxidized freestanding Si nanoparticles.\cite{carvalho,carvalho1} Moreover they agree excellently with the three-dimensional atomic probe tomography experimental data of Duguay and coworkers\cite{mathiot2012,khelifi2013} that demonstrated that n-type dopants are efficiently introduced in the "bulk" of the Si-NCs, whereas B atoms are preferentially located at their periphery, at the Si/SiO$_2$ interface.
Furthermore, we note from Figure~\ref{fig.totenergy} that in all the considered cases the doping of the SiO$_2$ region is unlikely to occur. In particular, for N and P doping is required a very high formation energy to move the dopant atom from the NC core to the silica (3.9 eV and 7.9 eV respectively), in agreement with the observation that for Si nanoclusters embedded in a SiO$_2$ matrix, the matrix provides a strong barrier to P diffusion inducing P segregation in the Si rich region.\cite{perego2012} Instead, still consistently with experimental results\cite{xie2013}, the diffusion of B and Al toward SiO$_2$ results significantly easier, with barriers of just 1.1 eV and 1.6 eV, respectively.
\\Actually, the presence of defects in real samples should foster interstitial doping in the silica and at the Si/SiO$_2$ interface. Such condition should be superposed to the above one in which substitutional doping acts on the basis of the lowest formation energy in defect-free samples.\cite{khelifi2013, gutsch2012}
\\It is worthwhile to note that a similar picture has been observed for doped Si nanowires (NWs). Fukata et al. \cite{fukata2013}, in a joint experimental-theoretical effort, have been shown that, whereas the impurity B atoms are more stable at the Si-SiO$_2$ interface in a split interstitial configuration at a Si site, P impurities preferentially piles up in the crystalline Si region.
\\The correspondence of the dopant positioning in different Si nanostructures, independently on their size or shape, indicates that the energetics is mainly governed by the atomic binding energy of the dopant, while other parameters (e.g. strain) play a minor role.
\\To better understand the effect of the different dopants on the NCs structural properties we have reported in Figure~\ref{fig.bond_lengths} the length of the bonds formed by the dopant with the four neighbouring atoms of the substituted Si.
We observe that P and Al maintain a sort of tetra-coordination in all the cases, with all the bonds showing approximately constant characteristic lengths with Si and O.
Instead, N tends to form in all the case a strong anti-bonding characteristics with one of the neigbours, that is repulsed at large distance. The same occurs for B, but only when placed at the interface with two oxygens (interface2), corresponding to the lowest-energy configuration.
\\To further investigate the local configuration at the impurity site, we provide in Table~\ref{table1} the charge population analysis of all the structures, obtained by the Mulliken method,\cite{mulliken} and the electronegativities of the considered atomic species.

Electronegativity plays a crucial role in determining the polarity of the bonds, i.e. the amount of charge localized onto a certain atom. It is interesting to note that the most stable sites correspond for the n-type dopants to the highest values in terms of population, while for the p-type dopants to the lowest ones. Due to the much higher electronegativity of oxygen with respect to silicon, the charge population of the dopant decrease when moving from the NC core to the silica. Such expectation is fulfilled in all cases except when the P and B dopants are localized in the silica. In our opinion this should be related to the proximity of the four neighboring oxygen atoms (see Fig.~\ref{fig.bond_lengths} top-right panel), leading to an overestimation of the charge population.

\begin{table}[t!]
  \begin{tabular}{l|@{\hspace{0.3cm}}c@{\hspace{0.3cm}}cc@{\hspace{0.2cm}}c@{\hspace{0.2cm}}}
    \multicolumn{5}{c}{Mulliken population} \\
    \hline
     Si$_{35}$/SiO$_2$ & core & interface1 & interface2 & silica \\
    \hline
     N                 & {\bf 5.38} & 5.12 & 5.00       & 4.83  \\
     P                 & {\bf 5.13} & 4.99 & 4.86       & 5.08  \\
     Al                & 3.73       & 3.27 & {\bf 2.84} & 2.57  \\
     B                 & 3.31       & 3.29 & {\bf 3.20} & 3.39  \\
    \hline
  \end{tabular}

  \rule{0pt}{2ex}

  \begin{tabular}{|@{\hspace{0.2cm}}c@{\hspace{0.2cm}}|@{\hspace{0.2cm}}c@{\hspace{0.2cm}}|@{\hspace{0.2cm}}c@{\hspace{0.2cm}}|@{\hspace{0.2cm}}c@{\hspace{0.2cm}}|}
    \multicolumn{4}{c}{Electronegativity} \\
    \hline
      B   &  C   &  N   &  O   \\
     2.04 & 2.55 & 3.04 & 3.44 \\
    \hline
      Al  &  Si  &  P   &  S   \\
     1.61 & 1.90 & 2.19 & 2.58 \\
    \hline
  \end{tabular}
  \caption{ Top: Mulliken population of the dopant atom for the Si$_{35}$ NC embedded in SiO$_2$ and doped with either N, P, Al, or B, substitutional of a Si atom in the NC center (core), at the interface bonded to one or two oxygens (interface1, interface2), or in the SiO$_2$ far from the NC (silica). Bold text highlights the most stable site for each dopant (see Fig.~\ref{fig.totenergy}). Bottom: Pauling electronegativities of the two upper lines of the periodic table, from group III to VI. }\label{table1}
\end{table}

\vspace{0.2cm}
\section{Co-doped Silicon Nanocrystals}\label{sec.co-doped}
Coming to the case of multiple dopant atoms, in particular to that of compensated co-doping, one should correspondingly increase the size of the hosting Si-NCs, in order to avoid a forced aggregation of the impurities. As already stated above, the energetics of the doping sites is mainly affected by the termination type at the interface, while the presence of an embedding matrix plays only a minor role. Therefore we have employed freestanding oxidized NCs as an approximation of the fully embedded ones, significantly reducing the computational effort and making possibile to consider larger NCs, that are more suitable in the presence of multiple dopant atoms. Moreover this choice allows us to take contact with the experiments performed on all-inorganic colloidal Si-NCs, co-doped with B and P, and terminated by OH.\cite{fujii2012,fujii2013a,fujii2013b}

\begin{figure*}[t!]
  \centering
  \includegraphics[draft=false,width=15cm]{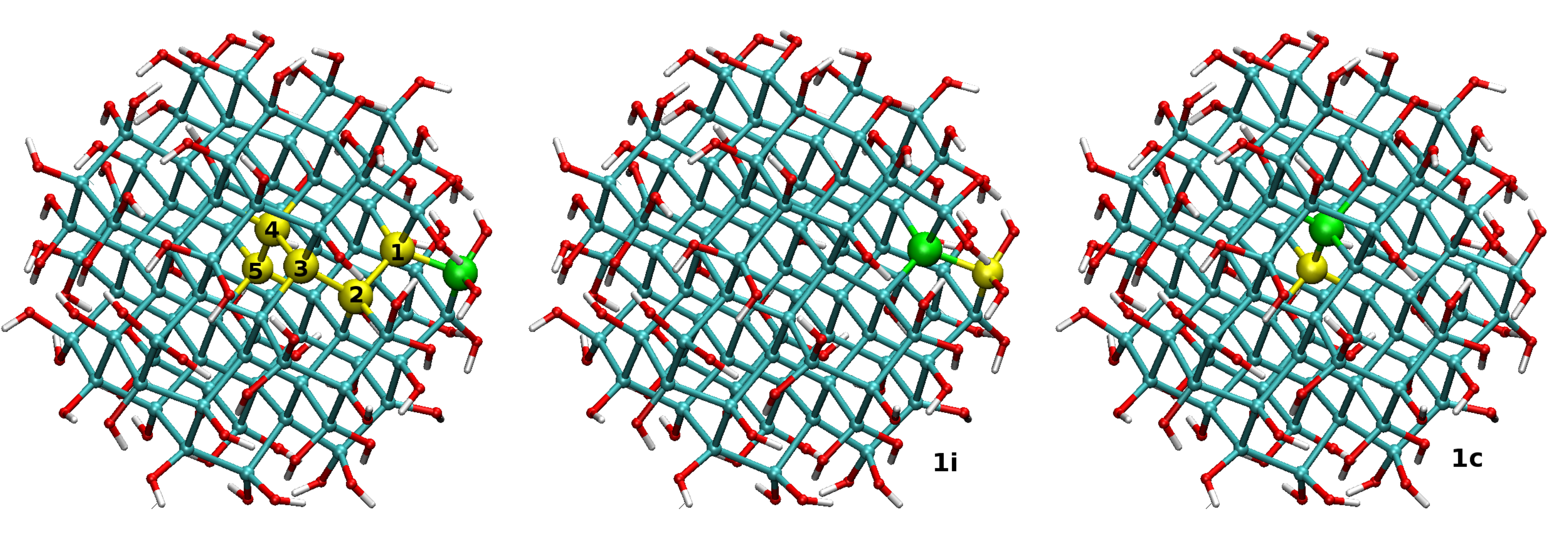}
  \caption{\small Freestanding OH-terminated Si$_{87}$ NC (left panel) doped with B at the surface and P placed at one of the sites from the center (5) to the nearest neighbour of B (1); (central panel) doped with P placed at the surface and B in the inner nearby (1i); (right panel) doped with P at the NC center and B in the outer nearby (1c). Cyan, red, white, green, and yellow spheres represent Si, O, H, B, and P atoms, respectively.}\label{fig.codoped_structures}
\end{figure*}

\begin{figure}[b!]
  \centering
  \includegraphics[draft=false,width=7.5cm]{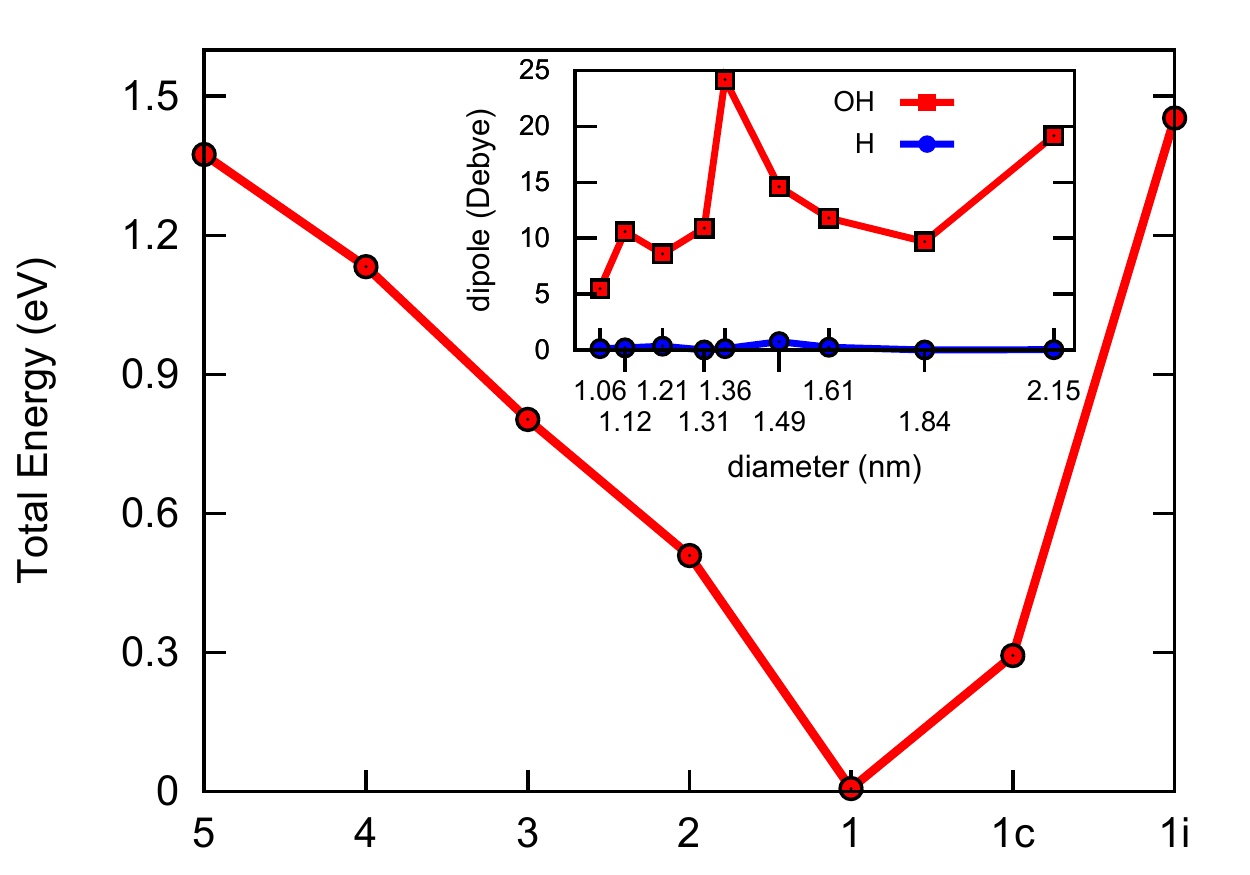}
\caption{\small Total energy of the co-doped freestanding Si$_{87}$-(OH)$_{76}$ as a function of the dopants position (see Fig.~\ref{fig.codoped_structures}). Inset: total dipole moment of the undoped NCs, terminated by H or OH groups, as a function of the NC size. }\label{fig.codoped_totenergy}
\end{figure}

The case of co-doped nanostructures presents interesting aspects related to the possibility of the dopant atoms to interact and bind together. In particular, in the case of opposite dopant types (i.e. n- and p-type), the emergence of new chemical pathways that can alter the picture obtained in the single-doped case has been proved for both Si NCs and NWs.\cite{fukata2013,ossicini1,ossicini2,ossicini3,ng2012}
\\To shed light on these aspect we have considered, as starting point, a freestanding oxidized NC, Si$_{87}$(OH)$_{76}$, co-doped with P and B atoms located at their energetically favored sites, as determined above for the single-doped case: P at the NC center and B at the NC surface. Further, to investigate the effect of the distance and of the interaction between the dopants we have considered additional nanostructures in which the P atom progressively approaches the B atom (see Fig.~\ref{fig.codoped_structures}, left panel). We have numbered the position of the P atom from 5 (center) to 1 (inner surface). From the analysis of the total energy, reported in Figure~\ref{fig.codoped_totenergy}, it is clear that while in the single-doped NC the P atom stabilize at the NC center, in the co-doped case the formation of a strong P-B bond (position 1) is clearly energetically favored, with a gain of about 1.4\,eV.
\\In alternative, the P-B bond can be formed at the NC center, fulfilling the energetics of the P dopant in the single-doped case while forcing the B atom in the NC core (see Fig.~\ref{fig.codoped_structures}, right panel). Such configuration, named 1c, results more energetically favored than the 5, 4, 3, and 2 ones, confirming the idea that the formation of the P-B bond rules over the other parameters (see Fig.~\ref{fig.codoped_totenergy}). Hovewer, the P-B pair at the interface (position 1) reveals more stable than that at the NC center (position 1c), with a difference in the formation energies of about 0.3\,eV .
\\Due to the polar nature of the P-B bond, the above condition implies the formation of a static electric dipole radially directed and preferentially located at the NC surface, pointing toward the NC center. The high stability of the dipole direction is evidenced by considering a configuration in which the co-dopants are switched, named 1i (see Fig.~\ref{fig.codoped_structures}, center panel). Quite interestingly, such configuration produce a total energy higher than all the other systems, and about 1.45\,eV higher than the most stable case.
\\We have computed the induced dipole as the difference between the total electric dipole of co-doped and undoped NCs. For the most stable configuration we have calculated an induced dipole moment $\mu \simeq 1$\,D, corresponding to two opposite charges of about 0.1\,a.u.\ placed at a distance of 2.32\,\AA~(the P-B bond length).
\\Despite the modest magnitude of the single dipole, the simultaneous presence of many inward-pointing dipoles shall be sustained by the strong stability of the dipole direction. In the case of the Si$_{87}$-(OH)$_{76}$ doped with six P-B pairs disposed along the three directions (two pairs for each axis), we have computed a difference in the formation energy of 28\,meV when switching one of the P-B pair atom positions (i.e. inverting the dipole direction), meaning that up to six inward-pointing dipoles are energetically permitted at environmental temperature at this NC size.
Besides, we expect that a much greater number of dipoles should be supported in a large NC, due to larger dipole-dipole internal separations (i.e. lower repulsive energies). The latter consideration is in agreement with Sugimoto and coworkers, reporting six dopant pairs in colloidal NCs of diameter 1.5\,nm, and 10-1000 dopant pairs in NCs of diameter 2-10\,nm, respectively.\cite{fujii2013b}

In a series of papers\cite{fujii2012,fujii2013a,fujii2013b}, Sugimoto and coworkers suggested that in the strong doping regime the co-doped NCs should be depicted by an internal shell positively charged surrounded by an external shell negatively charged, implying substantial repulsive forces between neighboring NCs that enable the stability of the colloidal Si-NCs in polar liquids without the formation of precipitates.
\\However, unless some charge is injected during the NCs preparation, we can restrict to the case of neutral NCs. In such case, since no electric field is emitted by two (homogeneous) concentric spherical shells of opposite charge, we would expect vanishing repulsive forces for large NCs in the strong co-doping regime (i.e. homogeneous charge limit).
\\We worth to stress that, differently than simple hydrogenated NCs, oxygenated NCs show a non-negligible ``macroscopic'' dipole due to a strong charge trapping at the surface oxygens.\cite{zhou} To make this clear, we have performed a total-dipole calculation on a set of fully-relaxed NCs of different size, H- and OH-terminated, reported in the inset of Figure~\ref{fig.codoped_totenergy}.
\\Therefore, while our calculations support the shell model proposed by Sugimoto et al., we have a different belief about the mechanism of stability. In our opinion, the surface dipoles form to the detriment of the
``macroscopic'' dipole of the NC, naturally existing in the undoped case.
As a consequence, while the undoped NCs aggregate by the ``macroscopic'' dipoles interaction, the neutralization of such dipoles by co-doping produce unpolarized NCs that cannot aggregate. Eventually, the screening provided by the polar liquid, and the thermal energy, may help in contrasting the aggregation induced by a residual polarization.
\\This view is supported by our case of NC with six co-dopants, presenting a total dipole about 7\% smaller with respect to the undoped case. Clearly, for larger NCs containing hundreds of co-dopants, we expect a much larger dipole-cancellation effect. Beside we also note that the undoped Si$_{87}$ NC shows the largest dipole of the set (see Fig.~\ref{fig.codoped_totenergy} inset), thus difficult to cancel out.

In the case of co-doped embedded NCs, the surface dipoles could in principle be aligned by an external field during the annealing, and would remain so afterwards due to the cage effect of the embedding matrix. If so, the possibility of forming a silicon-based ferroelectric semiconductor may constitute a convenient alternative to conventional materials made by toxic or rare elements.\cite{scott}
\vspace{0.2cm}

\section{Conclusions}\label{sec.conclusions}
By means of theoretical calculations we have substitutionally positioned a dopant atom inside or outside Si-NCs embedded in a SiO$_2$ matrix, revealing that whereas p-type (Al and B) dopants preferentially site at the NC interface, the n-type ones (N and P) are energetically favored in the Si-NC core, in agreement with recent experimental observations.\cite{mathiot2012,khelifi2013,perego2012}
The energetics of the systems as a function of the dopant position shows that SiO$_2$ forms a very large diffusion barrier for P and N, and a much reduced one for B and Al.
The analysis of the bond lengths and charge population around the dopant atom reveals a correlation between the electronegativity of the impurity and its preferential location, a strong indication that the presented results should be generalizable to other Si nanostructures like oxygenated NWs or Si/SiO$_2$ superlattices.
\\Following the results for the single-doped Si-NCs, in the case of n-type and p-type compensated co-doping we have investigated the most stable configuration for P+B impurities. We have revealed that the formation of a P-B bond is energetically favored, with B positioned at the NC surface and P at the inner nearby.
As a consequence, NC co-doping involves the spontaneous formation of static dipoles at the NC surface, very stable and radially directed inwards the NC. In our view, the role of these dipoles is that of disrupting the ``macroscopic'' dipole of the NC which is associated to aggregative forces in colloidal suspensions, and can provide an alternative explaination about the possibility of producing stable colloidal Si-NCs without any surface functionalization process.\cite{fukuda2011}
The presented outcomes can be very compelling for a large set of applications, from photovoltaics, \cite{conibeer2012,beard2007,loper2013,tang2011} to luminescence\cite{fujii2012}, bioimaging,\cite{erogbogbo2011} and nanomechanics,\cite{kemp2013} among others.

\vspace{0.3cm}
\section*{Acknowledgment}
Computational resources were made available by CINECA-ISCRA parallel computing initiative. We acknowledge financial support from the European Community's Seventh Framework Programme (FP7/2007-2013) under Grant No.\,245977.


\footnotesize
\begin{thebibliography}{}

\bibitem{daldosso2009} Daldosso, N.; Pavesi, L.;, {\it Laser \&  Photon. Rev.} {\bf 2009}, {\it 3}, 508.
\bibitem{blas2009} Per\'alvarez, M.; Barreto, J.; Carreras, J.; Morales, A.; Navarro-Urrios, D.; Lebour, Y. Dom\'inguez, C.; Garrido, B. {\it Nanotechnology} {\bf 2009}, {\it 20}, 405201.
\bibitem{cheng2011} Cheng K.-Y; Anthony R.; Kortshagen, U. R.; Holmes, R. J. {\it Nano Lett.},  {\bf 2011}, {\it 11}, 1952.
\bibitem{mastronardi2011} Mastronardi, M. L.; Maier-Flaig, F.; Faulkner, D.; Henderson, E. J.; K\"ubel, C.; Lemmer, U.; Ozin, G. A.  {\it Nano Lett.}, {\bf 2012}, {\it 12}, 337.
\bibitem{conibeer2012} Perez-Wurfln, I.; Ma, L.; Lin, D.; Hao, X.; Green, M. A.; Conibeer, G. {\it Sol. Energ. Mat. Sol. C.} {\bf 2012}, {\it 100}, 65.
\bibitem{beard2007} Beard, M. C.; Knutsen, K. P.; Yu, P.; Luther, J. M:; Song, Q.; Metzger, W. K.; Ellingson, R. J.; Nozik, A. J.  {\it Nano Lett.} {\bf 2007}, {\it 7}, 2506.
\bibitem{Liu2009} Liu, C.-Y.; Holman, Z. C.; Kortshagen, U. R. {\it Nano Lett.},  {\bf 2009}, {\it 9}, 449.
\bibitem{gregor2012} Trinh, M. T.; Limpens, R.; de Boer, W. D. A. M.; Schins, J. M.;
Siebbeles, L. D. A.; Gregorkiewicz, T. {\it Nature Photon.} {\bf 2012}, {\it 6}, 316.
\bibitem{ivan2012} Govoni, M.; Marri, I.; Ossicini, S. {\it Nature Photon.} {\bf 2012}, {\it 6}, 672.
\bibitem{loper2013} L\"oper, P.; Canino, M.; Lopez-Vidrier, J.; Schnabel, M.; Schindler, F.; Heinz, F.; Witzky, A.; Bellettato, M.; Allegrezza, M.; Hiller, D.; Hartel, A.; Gutsch, S.; Hern\`andez, S.; Guerra, R.; Ossicini, S.; Garrido, B.; Janz, S.; Zacharias, M. {\it Phys. Status Solidi A} {\bf 2013}, {\it 210}, 669.
\bibitem{erogbogbo2008} Erogbogbo, F.; Yong, K.-T.; Roy, I.; Xu, G.X.; Prasad, P. N.; Swihart, M. T. {\it ACS Nano},  {\bf 2008}, {\it 2}, 873.
\bibitem{erogbogbo2011} Erogbogbo, F.; Yong, K.-T.; Roy, I.; Hu, R.; Law, W.-C.; Zhao, W.; Ding, H.; Wu, F.; Kumar, R.; Swihart, M. T.; Prasad, P. N. {\it ACS Nano},  {\bf 2011}, {\it 5}, 413.
\bibitem{fujii2012} Sugimoto, H.; Fujii, M.; Imakita, K.; Hayashi, S.; Akamatsu, K.; {\it J. Phys. Chem. C} {\bf 2012}, {\it 116}, 17969.
\bibitem{law2010} Liu, Y.; Gibbs, M.; Puthussery, J.; Gaik, S.; Ihly, R.; Hillhouse, H.; Law, M.  {\it Nano Lett.},  {\bf 2010}, {\it 10}, 1960.
\bibitem{fukuda2011} Fukuda, M.; Fujii, M.; Sugimoto, H.; Imakita, K.; Hayashi, S {\it Opt. Lett.} {\bf 2011}, {\it 36}, 4026.
\bibitem{fujii2013a} Sugimoto, H.; Fujii, M.; Imakita, K.; Hayashi, S.; Akamatsu, K. {\it J. Phys. Chem. C} {\bf 2013}, {\it 117}, 6807.
\bibitem{fujii2013b} Sugimoto, H.; Fujii, M.; Imakita, K.; Hayashi, S.; Akamatsu, K. {\it J. Phys. Chem. C} {\bf 2013}, {\it 117}, 11850.
\bibitem{balberg} Balberg, I.; Savir, E.; Jedrzejewski, J.; Nassiopoulou, A. G.; Gardelis, S.; {\it Phys. Rev. B} {\bf 2007}, {\it 75}, 235329.
\bibitem{khria2012} Khriachtchev, L.; Ossicini, S.; Iacona, F.; Gourbilleau, F. {\it Intern. Journ. of Photoenergy} {\bf 2012}, 872576.
\bibitem{gutsch2012} Gutsch, S.; Hartel, A. M.; Hiller, D.; Zakharov, N.; Werner, P.; Zacharias, M. {\it Appl. Phys. Lett.} {\bf 2012}, {\it 100}, 233115.
\bibitem{mathiot2012} Mathiot, D.; Khelifi, R.; Muller, D.; Duguay, S. {\it Mater. Res. Soc. Symp. Proc.} {\bf 2012}, {\it 1455}, e-proc. DOI: 10.1557/opl.2012.1238
\bibitem{JAP2013} Guerra, R.; Cigarini, F.; Ossicini, S. {\it J. Appl. Phys.} {\bf 2013}, {\it 113}, 143505.
\bibitem{Norris} Norris, D. J.; Efors, A. L., Erwin, S. C. {\it Science} {\bf 2008}, {\it 319}, 1776.
\bibitem{khelifi2013} Khelifi, R. Mathiot, D.; Gupta, R.; Muller, D.; Roussel, M.; Duguay, S. {\it Appl. Phys. Lett.} {\bf 2013}, {\it 102}, 013116.
\bibitem{siesta1} Soler, J. M.; Artacho, E.; Gale, J. D.; Garc\'ia, A.; Junquera, J.; Ordej\'{o}n, P.; S\'anchez-Portal, D. {\it J. Phys. : Condens. Matter} {\bf 2002}, {\it 14}, 2745.
\bibitem{siesta2} Ordej\'{o}n, P.; Artacho, E.; Soler, J. M. {\it Phys. Rev. B} {\bf 1996}, {\it 53}, R10441.
\bibitem{PRB2} Guerra, R.; Degoli, E.; Ossicini, S. {\it Phys. Rev. B} {\bf 2009}, {\it 80}, 155332.
\bibitem{PRB3} Guerra, R.; Ossicini, S. {\it Phys. Rev. B} {\bf 2009}, {\it 81}, 245307.
\bibitem{kim2007} Kim, S.; Kim, M. C.; Choi, S.-H.; Kim, K. J.;  Hwang, H. N.; Hwang, C. C. {\it Appl. Phys. Lett.} {\bf 2007}, {\it 91}, 103113.
\bibitem{carvalho} Carvalho, A.; \"Oberg, S.; Barroso, M.; Rayson, M. J.;  Briddon, P. {\it Phys. Status Solidi A} {\bf 2012}, {\it 209}, 1847.
\bibitem{carvalho1} Carvalho, A.; \"Oberg, S.; Barroso, M.; Rayson, M. J.;  Briddon, P. {\it Phys. Status Solidi B} {\bf 2013}, {\it 250}, 1799.
\bibitem{mt} Martyna, G.J.; Tuckerman, M.E. {\it J.Chem.Phys.} {\bf 1999}, {\it 110}, 2810.
\bibitem{perego2012} Perego, M.; Seguini, G.; Fanciulli, M. {\it Surf. Interface Anal.} {\bf 2013}, {\it 45}, 386.
\bibitem{xie2013} Xie, M.; Li, D.; Cheng, L.; Wang, F.; Zhu, X.; Yang, D. {\it Appl. Phys. Lett.} {\bf 2013}, {\it 102}, 123108.
\bibitem{fukata2013} Fukata, N.; Kaminaga, J.; Takiguchi, R.; Rurali, R.; Dutta, M.; Murakami, K. {\it J. Phys. Chem. C} {\bf 2013}, {\it 117}, 20300.
\bibitem{mulliken} Mulliken, R. S. {\it J. Chem. Phys.} {\bf 1955}, {\it 23}, 1833.
\bibitem{ossicini1} Ossicini, S.; Degoli, E.; Iori, F.; Luppi, E.; Magri, R.; Cantele, G.; Trani, F.; Ninno, D.  {\it Appl. Phys. Lett.} {\bf 2005}, {\it 87}, 173120.
\bibitem{ossicini2} Iori, F.; Degoli, E; MAgri, R.; Marri, I.; Cantele, G.; Ninno, D.; Trani, F.; Ossicini, S. {\it Phys. Rev. B} {\bf 2007}, {\it 76}, 08532.
\bibitem{ossicini3} Iori, F.; Ossicini,  S. {\it Physica E} {\bf 2009}, {\it 41}, 939.
\bibitem{ng2012} Ng, M.-F.; Tong, S. W.  {\it Nano Lett.} {\bf 2012}, {\it 12}, 6133.
\bibitem{zhou} Zhou, Z.; Brus, L.; Friesner, R. {\it Nano Lett.} {\bf 2003}, {\it 3}, 163.
\bibitem{scott} Scott, J. F. {\it Science} {\bf 2007} {\it 315}, 954.
\bibitem{tang2011} Tang, J.; Kemp, K. W.; Hoogland, S.; Jeong, K. S.; Liu, H.; Levina, L.; Furukawa, M., Wang, X.; Debnath, R.; Cha, D.; Chou, K. W.; Fischer, A.; Amassian, A.; Asbury, J. B.; Sargent E. H. {\it Nat. Mater.} {\bf 2011}, {\it 10}, 765.
\bibitem{kemp2013} Kemp, B. A.; Whitney, J. G. {\it Appl. Phys. Lett.} {\bf 2013}, {\it 102}, 141605.

\end{thebibliography}
\end{document}